\documentclass[aps,pra,showpacs,twocolumn,superscriptaddress]{revtex4-1}
\usepackage{amsmath}
\usepackage{amssymb}
\usepackage{graphicx}
\usepackage{epstopdf}
\usepackage{times,txfonts}
\usepackage{easybmat}
\usepackage{mathtools}
\usepackage{subfigure}

\usepackage[bookmarks=false]{hyperref}
\hypersetup{colorlinks=true,citecolor=blue,linkcolor=blue,urlcolor=blue,pdfstartview=FitH,bookmarksopen=true}
\usepackage{color,soul}

\begin{document}
\preprint{APS/123-QED}

\vspace{3mm}

\title{Experimental demonstration of the criterion for the prepare-and-measure nonclassicality}
\author{Maolin Luo}
\email{These authors contributed equally to this work.}
\affiliation{School of Physics and State Key Laboratory of Optoelectronic Materials and Technologies, Sun Yat-sen University, Guangzhou 510000, China}
\author{Xiaoqian Zhang}
\email{These authors contributed equally to this work.}
\affiliation{School of Physics and State Key Laboratory of Optoelectronic Materials and Technologies, Sun Yat-sen University, Guangzhou 510000, China}
\author{Xiaoqi Zhou\footnotemark[2]}
\email{zhouxq8@mail.sysu.edu.cn}
\affiliation{School of Physics and State Key Laboratory of Optoelectronic Materials and Technologies, Sun Yat-sen University, Guangzhou 510000, China}
\affiliation{Hefei National Laboratory, University of Science and Technology of China, Hefei 230088, China}

\date{\today}

\begin{abstract}
  The prepare-and-measure theory is a new type of quantum paradox that reveals the incompatibility between classical theory and quantum mechanics in terms of the dimensionality of physical systems.
  Just as the Horodecki criterion can determine whether given quantum states are capable of exhibiting Bell nonclassicality, a similar criterion is needed for the prepare-and-measure theory to determine whether given quantum states can exhibit the prepare-and-measure nonclassicality.
  Recently, Poderini \emph{et al.} [Phys. Rev. Research 2, 043106 (2020)] presented such a criterion for the prepare-and-measure nonclassicality.
  In this work, we experimentally validate this criterion --- 52 different sets of quantum states are prepared and tested one by one using this criterion to determine whether they can exhibit the prepare-and-measure nonclassicality, and the experimental results are in good agreement with the theoretical expectations.
  The criterion experimentally verified here has the potential to be widely used in future research on the prepare-and-measure nonclassicality.
\end{abstract}

\maketitle

Since the introduction of the Einstein-Podolsky-Rosen paradox \cite{1EPR} in 1935, scholars have conducted in-depth studies on the nonclassicality of quantum systems, leading to the discovery of a series of nonclassical phenomena. These include quantum entanglement \cite{2Hor}, quantum steering \cite{3Uola}, Bell nonlocality \cite{2Bell,3nonlocal,4Clauser,5Chris,6Giu}, quantum contextuality \cite{7KS,8Mermin,9Cabello,10Zu,11Zhang}, and Leggett-Garg nonclassicality \cite{12Pal}. Quantum entanglement, quantum steering, and Bell nonlocality exhibit nonclassical behavior that conflicts with classical theories, which assume that particles have definite properties that can be measured independently of each other. In contrast, quantum contextuality and Leggett-Garg nonclassicality require only one particle to exhibit nonclassicality. These different nonclassical properties reveal contradictions between quantum and classical theories from different perspectives and are not only important in quantum foundations but also as resources for quantum information processing.
For example, quantum entanglement, quantum steering, and quantum nonlocality play essential roles in quantum communication \cite{13Ekert,14Bran,15ACn}, while quantum contextuality has important applications in oblivious transfer \cite{15Zava}, random number generation \cite{16Burn,17Fioren}, and quantum computing \cite{18Rauss,19Howard}. Therefore, researchers in the field of quantum information are highly motivated to explore possible new non-classical phenomena in quantum systems.

Recently, a new paradox in a prepare-and-measure (PAM) scenario \cite{12Gallego} has been proposed to demonstrate the incompatibility between quantum mechanics and classical models using only a two-dimensional quantum system. In this PAM paradox, one prepares a number of two-dimensional quantum states and then measures them accordingly to obtain results that violate an inequality that cannot be violated by the classical model.
In recent years, several theoretical \cite{13Bowles,14Sun} and experimental \cite{15Ahre,16Hend,17Ambr,18Lap} investigations have been reported based on the PAM paradox. In all of these works, specific quantum states need to be used to violate the PAM inequalities. Just as the Horodecki criterion \cite{19Horode} can be used to determine whether a quantum system has Bell nonlocality and the entanglement witness criterion \cite{19ghne,20zhu} can be used to determine whether a quantum system has quantum entanglement, Poderini et al. \cite{20Davide} proposed a criterion that can quickly determine whether given quantum states can exhibit PAM nonclassicality. In this Letter, we experimentally demonstrated this criterion using 52 sets of single-qubit pure states and the experimental results agree well with the theoretical predictions. Future research on the PAM nonclassicality could make extensive use of the criterion experimentally verified here.

Before presenting our experiments, let us briefly review the standard PAM theory.
A PAM scenario comprises a preparation device $P$ and a measurement device $M$ that receive random variables $j$ and $k$ as inputs, respectively. The preparation device $P$ prepares different physical systems according to the value of $j$ and sends them to the measurement device $M$.
The measurement device $M$ performs different measurements of the received physical system according to the value of $k$. The resulting outcome is $b$, and its probability distribution is given by $p(b|j,k)$.
In classical descriptions, the physical systems being prepared can be represented by a random variable $s$, which is not empirically observable. As the choice of which observable to measure is made after the system has been prepared, $s$ does not depend on $k$. Consequently, the probability distribution of the measurement result $b$ can be expressed as
\begin{eqnarray}
p(b|j,k)=\sum\limits_{s}p(b|s,k)p(s|j).
\end{eqnarray}

\begin{figure}[!ht]
  \centering
  \includegraphics[width=0.35\textwidth]{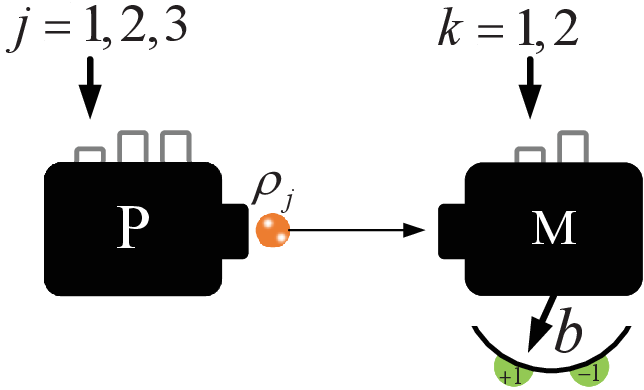}
  \caption{(a) The prepare-and-measure scenario. $P$ denotes the state preparation device and $M$ denotes the measurement device.
  When button $j$ is pressed, the state preparation device sends the $j$th state to the measurement device. When button $k$ is pressed, the measurement device performs measurement $M_k$ on this state and produces the outcome $b\in\{+1,-1\}$.}\label{fig2}
\end{figure}

In classical descriptions, any randomness present in the probabilities can be absorbed into a set of hidden variables $\lambda$, yielding
\begin{eqnarray}
p(b|j,k)=\sum\limits_{\lambda,s}p(b|s,k,\lambda)p(s|j,\lambda)p(\lambda).
\end{eqnarray}

We now turn our attention to a quantum description of the PAM scenario.
In the quantum description, the preparation device P generates a quantum state $\rho_j$ based on the value of j and sends it to the measurement device $M$. The measurement device $M$ then measures the observable $M^k$ based on the value of $k$.
According to Born's law, the probability distribution of the measurement result $b$ is given by
\begin{eqnarray}
p(b|j,k) = Tr(\rho_jM^k).
\end{eqnarray}

From the above derivation, it is evident that classical and quantum theories describe the probability distribution of the measurement result $b$ differently. If the probability distribution of $b$ in a particular PAM experiment cannot be explained by the classical description, it indicates the existence of PAM nonclassicality in the physical system used in that experiment.

Let us consider the simplest scenario below.
As shown in Fig. 1, there is a state preparation device, which has three buttons corresponding to three two-dimensional physical states, and a state measurement device, which has two buttons corresponding to each of the two measurements.

Before the start of each round of PAM, one of the three buttons of the preparation device ($j$th) and one of the two buttons of the measurement device ($k$th) are randomly chosen and pressed.
The preparation device then sends the $j$th physical state to the measurement device, which then performs the $k$th measurement on it, producing the measurement result $b\in\{+1,-1\}$. If the above physical process of the PAM can be described classically, one can obtain the following inequality \cite{12Gallego}
\begin{eqnarray}
S=E_{11}+E_{12}+E_{21}-E_{22}-E_{31}\leq3,\label{eq1}
\end{eqnarray}

where $E_{jk}=p(+1|j,k)-p(-1|j,k)$ denotes the expected value and $p(b|j,k)$ denotes the probability of obtaining the outcome $b$ when the $k$th measurement is performed on the $j$th state.
However, if the preparation device prepares three quantum states, then the maximum value of S obtained can reach as high as $1+2\sqrt{2}$, thus violating the above inequality and revealing the incompatibility of the classical and quantum descriptions.

Although the above inequality can be violated using quantum resources, not all quantum states can achieve the violation of this PAM inequality. Therefore, it is necessary to find a criterion that can determine whether given quantum states can violate the inequality. Recently, Poderini \emph{et al.} \cite{20Davide} have found a criterion to quickly determine whether the given three two-dimensional quantum states can violate the PAM inequality or not.
We will briefly explain how this criterion is obtained below.
As shown in Fig. 1, the quantum states produced by the preparation device can be expressed as
\begin{eqnarray}
\rho_j=1/2(I+\overrightarrow{r_j}\cdot \overrightarrow{\sigma})\quad (j=1,2,3)\label{eq2}
\end{eqnarray}
where $\overrightarrow{\sigma}=(\sigma _x,\sigma_y,\sigma_z)$ ($\sigma_x,\sigma_y,\sigma_z$ are Pauli operators).
The projection operators corresponding to the measurements are $M_k^{\pm}=1/2(I\pm \overrightarrow{q_k}\cdot \overrightarrow{\sigma})\quad (k=1,2)$, where $\overrightarrow{q_k}$ is a Bloch vector and $||\overrightarrow{q_k}||=1$.
By calculation, one can get
\begin{eqnarray}
\begin{split}
E_{jk}&=p(+1|j,k)-p(-1|j,k)\\
&=Tr(\rho_jM_k^{+})-Tr(\rho_jM_k^-)=\overrightarrow{r_j}\cdot\overrightarrow{q_k}.\label{eq3}
\end{split}
\end{eqnarray}
Substituting Eq. \eqref{eq3} into Eq. \eqref{eq1}, one gets
\begin{eqnarray}
S=\overrightarrow{q_1}\cdot (\overrightarrow{r_1}+\overrightarrow{r_2}-\overrightarrow{r_3})+\overrightarrow{q_2}\cdot(\overrightarrow{r_1}-\overrightarrow{r_2}).\label{eq8}
\end{eqnarray}

When all three prepared states are pure, the relationship between $\overrightarrow{r_1}$, $\overrightarrow{r_2}$ and $\overrightarrow{r_3}$ can be described by $\theta_1$, $\theta_2$ and $\theta_3$, where
\begin{eqnarray}
\overrightarrow{r_1}\cdot \overrightarrow{r_2}=cos2\theta_1,\quad \overrightarrow{r_2} \cdot \overrightarrow{r_3}=cos2\theta_2, \quad \overrightarrow{r_1}\cdot \overrightarrow{r_3}=cos2\theta_3.\label{eq5}
\end{eqnarray}
It is clear that the maximum value of S
\begin{eqnarray}
\begin{split}
S_{max}&=||\overrightarrow{r_1}+\overrightarrow{r_2}-\overrightarrow{r_3}||+||\overrightarrow{r_1}-\overrightarrow{r_2}||\\
&=\sqrt{3+2cos2\theta_1-2cos2\theta_2-2cos2\theta_3}+2sin\theta_1
\end{split}\label{eq6}
\end{eqnarray}
can be obtained when $\overrightarrow{q_1}$ and $\overrightarrow{q_2}$ are set to $\frac{\overrightarrow{r_1}+\overrightarrow{r_2}-\overrightarrow{r_3}}{||\overrightarrow{r_1}+\overrightarrow{r_2}-\overrightarrow{r_3}||}$ and $\frac{\overrightarrow{r_1}-\overrightarrow{r_2}}{||\overrightarrow{r_1}-\overrightarrow{r_2}||}$, respectively.
Since the three prepared states can be relabeled, there are two other maximum values
\begin{eqnarray}
\begin{split}
S'_{max}&=\sqrt{3+2cos2\theta_3-2cos2\theta_2-2cos2\theta_1}+2sin\theta_3,\\
S''_{max}&=\sqrt{3+2cos2\theta_2-2cos2\theta_1-2cos2\theta_3}+2sin\theta_2.
\end{split}\label{eq7}
\end{eqnarray}
Putting it all together, any one of the three values of $S_{max}$, $S'_{max}$, and $S''_{max}$ being greater than 3 would indicate that the given three quantum states can violate the PAM inequality.

To explain more clearly how we determine the measurement bases and whether the PAM inequality is violated for given input states, we provide a concrete example.
Suppose the three quantum pure states are $\rho_1 = |+\rangle\langle+|$, $\rho_2 = |0\rangle\langle0|$ and $\rho_3 = |\phi\rangle\langle\phi|$, where
$|+\rangle=\frac{1}{\sqrt{2}}(1, 1)^T$, $|0\rangle=(1, 0)^T$ and $|\phi\rangle=(-0.2588i, 0.9659)^T$. According to Eq. \eqref{eq2}, the corresponding three vectors can be calculated as $\overrightarrow{r_1}=[1,0,0]$, $\overrightarrow{r_2}=[0,0,1]$ and $\overrightarrow{r_3}=[0,\frac{1}{2},-\frac{\sqrt{3}}{2}]$, respectively.
The angles between them can be calculated as $\theta_1=\frac{3\pi}{12}$, $\theta_2=\frac{5\pi}{12}$ and $\theta_3=\frac{\pi}{4}$ according to Eq. \eqref{eq5}.
Substituting the values of these three angles into Eq. \eqref{eq6} and Eq. \eqref{eq7}, one can obtain $S_{max}=3.5895$, $S'_{max}=3.5895$ and $S''_{max}=3.0579$.
By selecting the measurement bases $\{|\psi_1^+\rangle, |\psi_1^-\rangle\}$, $\{|\psi_2^+\rangle, |\psi_2^-\rangle\}$, where $|\psi_1^+\rangle=(0.8620+0.4310i, 0.2666)^T$, $|\psi_1^-\rangle=(-0.2666, 0.8620-0.4310i)^T$, $|\psi_2^+\rangle=(0.3827, 0.9239)^T$ and $|\psi_2^-\rangle=(-0.9239, 0.3827)^T$, the maximum S value of 3.5895 can be obtained, thus confirming that this set of quantum states can violate the PAM inequality.

We now present the experiment for validating the PAM criterion. As shown in Fig. 2, a 50 mW ultraviolet laser with a central wavelength of 405 nm is focused on a beta-barium borate (BBO) crystal to produce a pair of photons via type II spontaneous parametric down-conversion.
After one of the photons is triggered, the other photon is prepared to a specific single-qubit pure state ($\rho_1$, $\rho_2$ or $\rho_3$) via a half-wave plate (HWP1) and a quarter-wave plate (QWP1) and sent to the measurement device.
The measurement device consisting of QWP2, HWP2, the polarization beamsplitter (PBS1) and the single photon counting modules (SPCMs) measures the quantum states at specific bases ($|\psi_1^\pm\rangle$ or $|\psi_2^\pm\rangle$), thus completing the PAM process.

\begin{figure}[!h]
  \flushleft
  \includegraphics[width=0.35\textwidth]{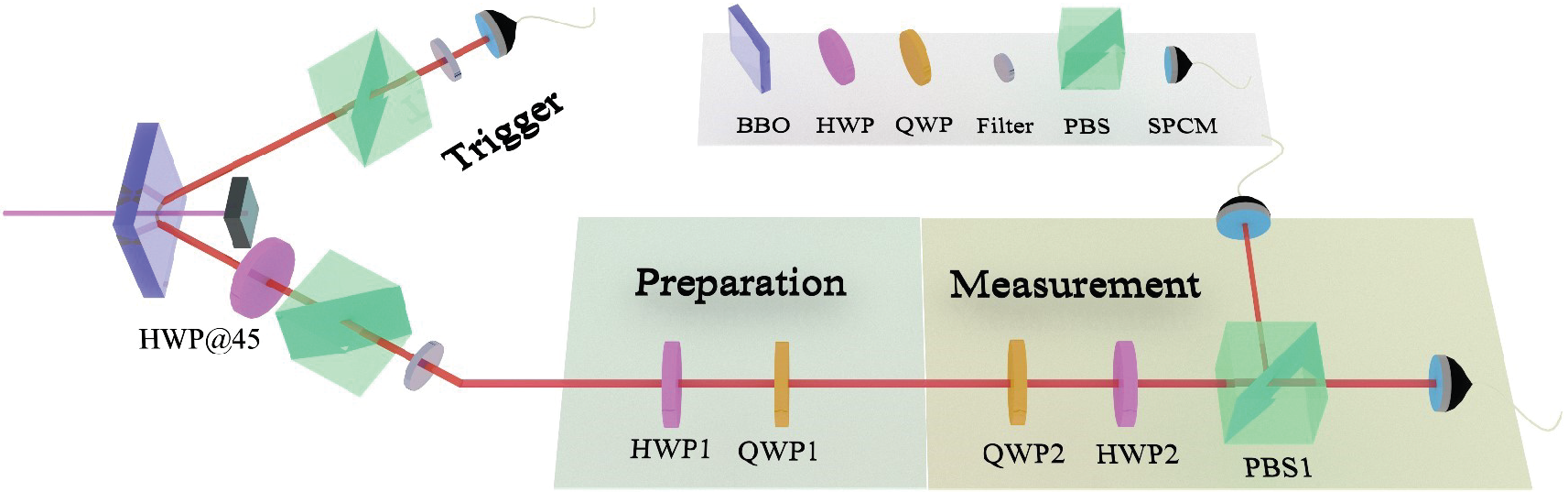}
  \caption{Experimental setup for the prepare-and-measure scenario. A 50 mW continuous ultraviolet laser with a central wavelength of 405 nm passes through a type-II beta-barium borate (BBO) crystal to produce a photon pair.
  One of the two photons is triggered and the other photon is prepared to the desired quantum state through a half-wave plate (HWP1) and a quarter-wave plate (QWP1).
  The prepared quantum state is then measured by a measurement device consisting of QWP2, HWP2, the polarization beamsplitter (PBS1) and two single photon counting modules (SPCMs) on the desired bases.  }\label{fig2}
\end{figure}

\begin{figure*}[!htb]
  \begin{minipage}[t]{1\textwidth}
  \centering
  \includegraphics[width=0.43\textwidth]{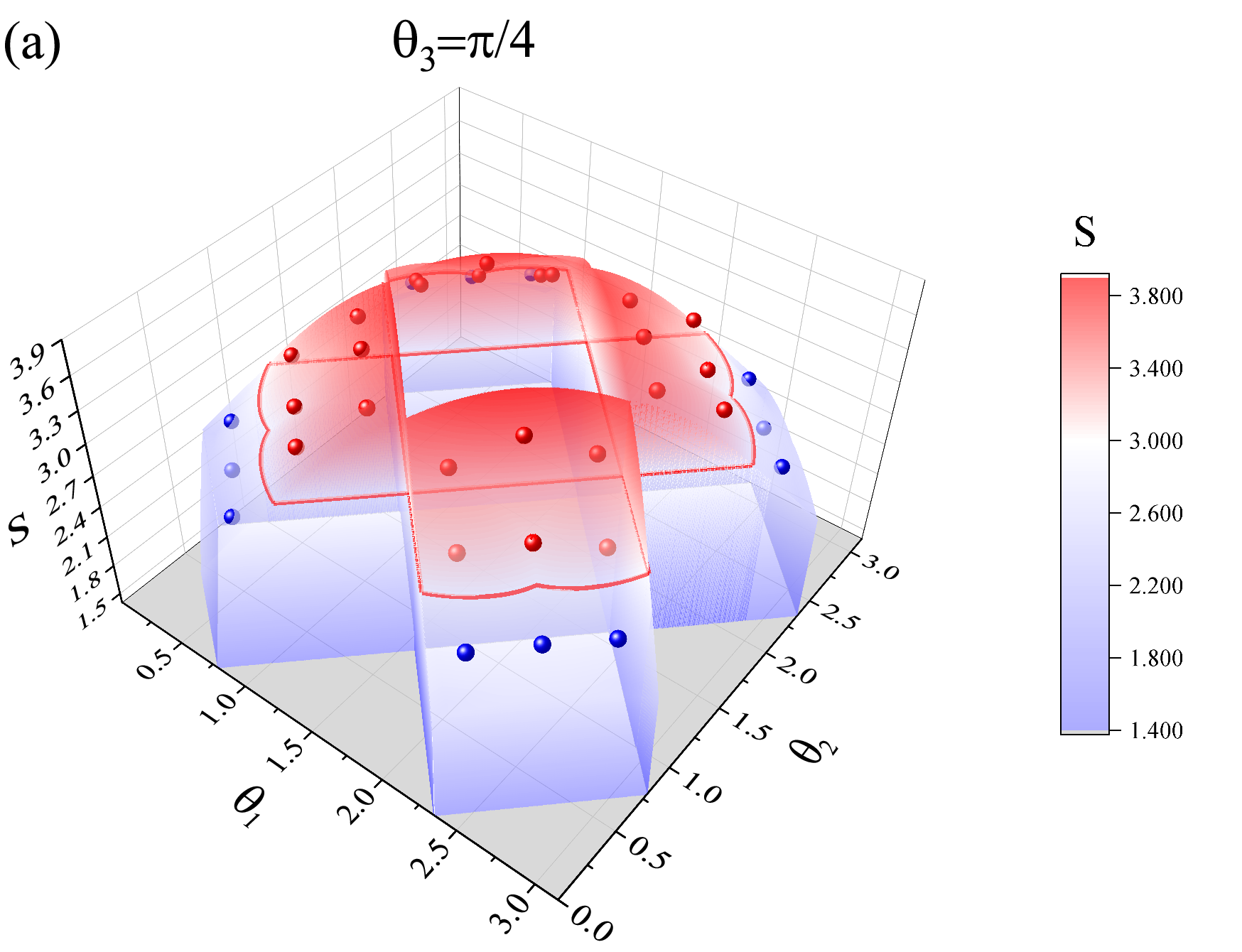}\qquad
  \includegraphics[width=0.43\textwidth]{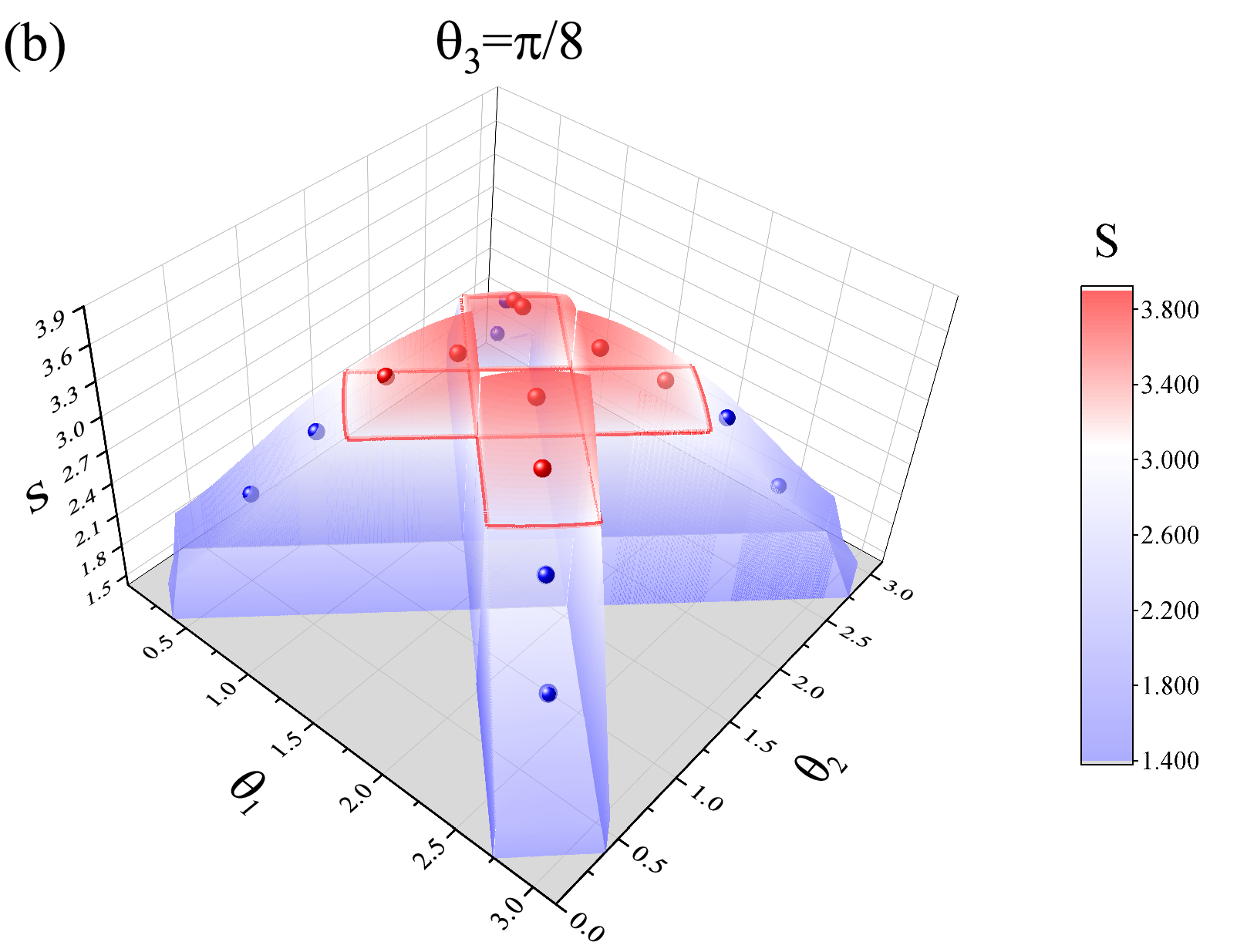}
  \end{minipage}
  \caption{Experimental results for testing the prepare-and-measure (PAM) criterion. (a) and (b) correspond to the experimental results for 36 sets of quantum states with $\theta_3=\frac{\pi}{4}$ and 16 sets of quantum states with $\theta_3=\frac{\pi}{8}$, respectively. The red area is the region where the theory predicts that the PAM inequality can be violated, and the blue area is the region where the theory predicts that the PAM inequality cannot be violated. The gray area corresponds to the region that do not exist physically.  }\label{fig3}
\end{figure*}

To test this PAM criterion, in our experiment we selected 52 sets of single-qubit pure states $\{\rho_1, \rho_2, \rho_3\}$ to be sent to the measurement device, of which 36 sets of quantum states have $\theta_3$ as $\frac{\pi}{4}$ and 16 sets of quantum states have $\theta_3$ as $\frac{\pi}{8}$. The values of $\theta_1$ and $\theta_2$ for these 52 sets of quantum states cover a variety of cases.
Figure 3a shows the experimental results for the 36 sets of quantum states with $\theta_3=\frac{\pi}{4}$ and Fig. 3b shows the experimental results for the 16 sets of quantum states with $\theta_3=\frac{\pi}{8}$.
In our experiment, we repeated the measurements 50 times for each transmitted quantum state to calculate the error bar of S. Each measurement had an acquisition time of 2 seconds, with 1 second for each of the two sets of measurement bases. The count rate per second was approximately 2500, which is sufficient to mitigate the potential errors caused by Poisson noise.
It can be seen that the measured values of S are in high agreement with the theoretical ones---the average deviation of the measured values from the theoretical values is about 2\%.
As predicted by the PAM criterion, when the given quantum state is in the red region, the S value would be greater than 3, and the PAM inequality can be violated; when the given quantum state is in the blue region, the S value would be less than 3, and the PAM inequality cannot be violated.

Note that, for pure states that cannot violate the PAM inequality with the above criterion, they are still not capable of violating the PAM inequality even if the number of measurements is increased.
However, for special mixed states, if the PAM inequality cannot be violated using two measurements, it is possible to violate the PAM inequality by increasing the number of measurements to three or more.
We have performed an experiment to investigate this phenomenon, and the details are given in Appendix A.

In summary, we experimentally validate the criterion for quickly determining whether given three quantum states can violate the prepare-and-measure inequality. We experimentally prepared 52 sets of quantum pure states and test them using this criterion, and the experimental results are in good agreement with the theoretical expectations.
The criterion experimentally verified here has the potential to be widely used in future research on the prepare-and-measure nonclassicality.

This work was supported by the National Natural Science Foundation of China (Grant No. 61974168), the National Key Research and Development Program (Grant No. 2017YFA0305200) and the Key Research and Development Program of Guangdong Province of China (Grants No. 2018B030329001 and No. 2018B030325001). X. -Q. Zhou acknowledges support from Hefei National Laboratory. X.-Q. Zhang acknowledges support from the National Natural Science Foundation of China (Grant No. 62005321) and Natural Science Foundation of Guangdong Province of China (Grant No. 2023A1515011556). M. -L. Luo acknowledges support from the Guangdong Basic and Applied Basic Research Foundation (Grant No. 2019A1515011048).

\section*{APPENDIX A: Activation of the prepare-and-measure nonclassicality}
In the PAM scenario, one can observe the following interesting phenomenon.
For a given set of quantum mixed states, the PAM nonclassicality cannot be revealed when only two sets of measurement bases are used, but when an additional set of measurement basis is added, the PAM nonclassicality can then be activated.
We have performed an experiment to demonstrate this phenomenon.

First, we prepared the following three mixed states $\rho_1$, $\rho_2$ and $\rho_3$, where

\begin{eqnarray*}
\begin{array}{l}
\displaystyle \rho_1=0.0981|\varphi_{1a}\rangle\langle\varphi_{1a}|+0.9019|\varphi_{1b}\rangle\langle \varphi_{1b}|,\\
\displaystyle \rho_2=0.0981|\varphi_{2a}\rangle\langle\varphi_{2a}|+0.9019|\varphi_{2b}\rangle\langle \varphi_{2b}|,\\
\displaystyle \rho_3=0.0981|\varphi_{3a}\rangle\langle\varphi_{3a}|+0.9019|\varphi_{3b}\rangle\langle \varphi_{3b}|,
\end{array}
\end{eqnarray*}
in which the six pure states $|\varphi_{1a}\rangle$, $|\varphi_{1b}\rangle$, $|\varphi_{2a}\rangle$, $|\varphi_{2b}\rangle$, $|\varphi_{3a}\rangle$ and $|\varphi_{3b}\rangle$ are of the following form
\begin{eqnarray*}
\begin{array}{l}
\displaystyle |\varphi_{1a}\rangle=0.2588|0\rangle-0.9659|1\rangle, \ |\varphi_{1b}\rangle=-0.9659|0\rangle-0.2588|1\rangle,\\
\displaystyle |\varphi_{2a}\rangle=-0.9659|0\rangle+0.2588|1\rangle, \ |\varphi_{2b}\rangle=0.2588|0\rangle+0.9659|1\rangle,\\
\displaystyle |\varphi_{3a}\rangle=-0.7071|0\rangle-0.7071|1\rangle, \ |\varphi_{3b}\rangle=-0.7071|0\rangle+0.7071|1\rangle.
\end{array}
\end{eqnarray*}
For the set of quantum states $\rho_1$, $\rho_2$ and $\rho_3$, the maximum $S$ value that can be obtained theoretically with only two sets of measurement bases is 2.7847, which is smaller than the classical bound 3.
We have experimentally chosen the two bases $\{|\psi_1^+\rangle, |\psi_1^-\rangle\}$ and $\{|\psi_2^+\rangle, |\psi_2^-\rangle\}$, where $|\psi_1^+\rangle=(cos\frac{\pi}{6}, sin\frac{\pi}{6})^T$, $|\psi_1^-\rangle=(sin\frac{\pi}{6}, -cos\frac{\pi}{6})^T$, $|\psi_2^+\rangle=(1, 0)^T$ and $|\psi_2^-\rangle=(0, 1)^T$.
The measured S value is $2.7123\pm0.024$, which indeed cannot violate the PAM inequality.
When an additional set of measurement basis is added, i.e., in the case of three quantum states and three measurement bases, the PAM inequality becomes $T=|E_{11}+E_{12}-E_{22}+E_{23}-E_{31}-E_{33}|\leq4$.
For the set of quantum states $\rho_1$, $\rho_2$ and $\rho_3$, the maximum T value that can be obtained theoretically is 4.1769, which is larger than the classical bound 4.
We have experimentally chosen the three bases $\{|\psi_1^+\rangle, |\psi_1^-\rangle\}$, $\{|\psi_2^+\rangle, |\psi_2^-\rangle\}$ and $\{|\psi_3^+\rangle, |\psi_3^-\rangle\}$, where $|\psi_3^+\rangle=(cos\frac{\pi}{3}, sin\frac{\pi}{3})^T$ and $|\psi_3^-\rangle=(sin\frac{\pi}{3}, -cos\frac{\pi}{3})^T$.
The measured value of $T$ is $4.1589\pm0.021$, thus verifying the activation of the PAM nonclassicality by adding an additional measurement basis.


%


\end{document}